\documentclass{article}
\usepackage{amsmath}
\usepackage{amssymb,amsfonts}
\usepackage[square,comma]{natbib}
\begin{document}
\begin{center}
\LARGE{\textsf{Compact anisotropic spheres with prescribed energy
density}}
\end{center}

\begin{center}
M. Chais$\mbox{i}^{\mbox{\scriptsize
1,2}}$\footnotetext[1]{Astrophysics and Cosmology Research Unit,
School of Mathematical Sciences, University of KwaZulu-Natal,
Durban 4041, South Africa. eMail:
\texttt{maharaj@ukzn.ac.za}}\footnotetext[2]{Department of
Mathematics \& Computer Science, National University of Lesotho,
Roma 180, Lesotho. eMail: \texttt{m.chaisi@nul.ls}} and S. D.
Mahara$\mbox{j}^{\mbox{\scriptsize 1}}$
\end{center}

\section*{Abstract}

New exact interior solutions to the Einstein field equations for
anisotropic spheres are found. We utilise a procedure that
necessitates a choice for the energy density and the radial
pressure. This class contains the constant density model of
Maharaj and Maartens (\emph{Gen. Rel. Grav.}, \textbf{21},
899-905, 1989)\nocite{MaharajMaartens} and the variable density
model of Gokhroo and Mehra (\emph{Gen. Rel. Grav.}, \textbf{26},
75-84, 1994)\nocite{gokhroo} as special cases. These anisotropic
spheres match smoothly to the Schwarzschild exterior and
gravitational potentials are well behaved in the interior. A
graphical analysis of the matter variables is performed which
points to a physically reasonable matter distribution.

\textbf{Keywords}: anisotropic relativistic stars; compact spheres

\section{Introduction}

In recent years a number of authors have studied solutions to the
Einstein field equations corresponding to anisotropic matter where
the radial component of the pressure differs from the angular
component. The gravitational field is taken to be spherically
symmetric and static since these solutions may be applied to
relativistic stars. A number of researchers have examined how
anisotropic matter affects the critical mass, critical surface
redshift and stability of highly compact bodies. These
investigations are contained in the papers
\cite{BowersLiang,DevGleiser,DevGleiser2,HerreraRugg,Ivanov,
MakHarko2002,MakHarko,Rago}, among others. Some researchers have
suggested that anisotropy may be important in understanding the
gravitational behaviour of boson stars and the role of strange
matter with densities higher than neutron stars. Mak and Harko
\cite{MakHarko2002} and Sharma and Mukherjee
\cite{SharmaMukherjee2002} suggest that anisotropy is a crucial
ingredient in the description of dense stars with strange matter.

In this paper our objective is to generate a new class of exact
solutions to the Einstein field equations corresponding to a
physically reasonable form for the energy density. A particular
motive is to find simple analytic forms for the gravitational and
matter variables so that the physical interpretation of the model
is simplified. Often the solutions are presented in terms of
special functions or a numerical approach is required
\cite{DevGleiser,DevGleiser2}. We hope that our results in terms
of elementary functions will assist in the analysis of
gravitational behaviour of compact objects, and the study of
anisotropy under strong gravitational fields. In section
\ref{sec:II} we develop the anisotropic stellar model and present
the relevant field equations. A particular form for the energy
density is chosen in section \ref{sec:III}, and the Einstein field
equations are integrated. Special cases of physical interest are
isolated from the general solution in section \ref{sec:IV}. Some
physical features of the anisotropic star are briefly considered
in section \ref{sec:V}. In section \ref{sec:VI} we demonstrate
that our model yields surface redshifts and masses that correspond
to real sources, and make a few concluding remarks and suggestions
for future research.
\section{The anisotropic model\label{sec:II}}

The line element for static spherically symmetric spacetimes is
given by
\begin{eqnarray}
\mbox{d}s^2 & = &
-e^{\nu}\mbox{d}t^2+e^{\lambda}\mbox{d}r^2+r^2\left(\mbox{d}\theta^2+\sin^2\theta\mbox{d}\phi^2\right)
\label{metric2}
\end{eqnarray}
where $\nu(r)$ and $\lambda(r)$ are arbitrary functions. We study
non-radiating relativistic spheres with anisotropic stress, and
the energy-momentum tensor is of the form
\begin{eqnarray}
T^{ab}& = & \mu u^au^b+ph^{ab}+\pi^{ab} \label{massT3}
\end{eqnarray}
where $\mu$ is the energy density, $p$ the isotropic pressure, and
the projection tensor $h^{ab}=u^au^b+g^{ab}$ is measured relative
to the four-velocity $u^a$. It is convenient to express the
anisotropic stress in the form
\begin{eqnarray*}
\pi^{ab} & = & \sqrt{3}S(r)\left(c^ac^b-\frac{1}{3}h^{ab}\right)
\end{eqnarray*}
where the unit spacelike vector $\mathbf{c}$ is orthogonal to the
fluid four-velocity $\mathbf{u}$ and $|S(r)|$ is the magnitude of
the stress tensor. This representation for $\pi^{ab}$ is a
consequence of the symmetries of the static spherically symmetric
spacetimes \cite{MaharajMaartens2}. The quantity $S$ is a useful
device to introduce
\[
p_r  =  p+2S/\sqrt{3}, \;\; p_\perp  =  p-S/\sqrt{3}
\]
which are the radial and tangential pressures respectively. Note
that for isotropic matter $S=0$ and $p_r=p_\perp=p$. The magnitude
$S$ provides a measure of anisotropy. We assume that the fluid
four-velocity is comoving. This assumption implies that $ u^a  =
e^{-\nu/2}\delta^a_0, \;\; c^a  = e^{-\lambda/2}\delta^a_1 $.

Using \eqref{metric2} and \eqref{massT3}, the Einstein field
equations become
\begin{subequations}
\begin{eqnarray}
-\frac{e^{-\lambda}}{r^2}\left(1-\lambda^\prime
r-e^\lambda \right)  & = & \mu \label{EFEs:1} \\
\frac{e^{-\lambda}}{r^2}\left(1-e^\lambda+r\nu^\prime\right)
& = & p_r \label{EFEs:2} \\
\frac{e^{-\lambda}}{4}\left(2\nu^{\prime\prime}-\nu^\prime\lambda^\prime
+{\nu^\prime}^2+\frac{2\nu^\prime}{r}-\frac{2\lambda^\prime}{r}\right)
& = & p_\perp \label{EFEs:3}
\end{eqnarray}
\label{EFEs}
\end{subequations}
for static spherically symmetric anisotropic matter. We are using
units where the speed of light and the coupling constant are
unity. The momentum conservation equation leads to
\begin{eqnarray}
\left(\mu+p_r\right)\nu^\prime+2p^\prime_r+\frac{4}{r}\left(p_r-p_\perp\right)
& = & 0 \label{massTdivEqn}
\end{eqnarray}
for the spacetime (\ref{metric2}). This conservation equation is
not independent and can be generated directly from the field
equations \eqref{EFEs}. We define the mass function as
\begin{eqnarray}
m(r) & = & \frac{1}{2}\int^r_0x^2\mu(x)\mbox{d}x \label{massFun}
\end{eqnarray}
following the treatment of Stephani \cite{stephani}. With the help
of (\ref{massTdivEqn}) and (\ref{massFun}) we can integrate
\eqref{EFEs:1}, and then get the equivalent system
\begin{subequations}
\begin{eqnarray}
e^{-\lambda} & = & 1-\frac{2m}{r} \label{EFEs2:1} \\
r(r-2m)\nu^\prime & = & p_r r^3+2m \label{EFEs2:2} \\
\left(\mu+p_r\right)\nu^\prime+2p^\prime_r & = &
-\frac{4}{r}\left(p_r-p_\perp\right) \label{EFEs2:3}
\end{eqnarray}
\label{EFEs2}
\end{subequations}
The system (\ref{EFEs2}) has the advantage of being a first order
system of differential equations, and is linear in the
gravitational potential $\nu$ which simplifies the integration
process. For certain applications it is easier to use
(\ref{EFEs2}) rather than the original second order system
(\ref{EFEs}), which is the approach that we follow in this paper.
We seek explicit solutions to the Einstein field equations that
describe realistic anisotropic relativistic stars by utilising an
algorithm that was initially proposed by Maharaj and Maartens
\cite{MaharajMaartens}. In their approach they expressed the field
equations as the first order system of differential equations
\eqref{EFEs2}. The energy density $\mu$ (or equivalently $m$) and
the radial pressure $p_r$ are chosen on physical grounds. The
remaining relevant quantities $(e^\nu,e^\lambda,p_\perp)$ then
follow from the field equations. Note that $(e^\nu,e^\lambda,\mu\;
(\mbox{or}\; m),p_r,p_\perp)$ are not independent; there are five
unknown functions and three field equations so that we have the
freedom to choose any two of the quantities. In this paper we make
explicit choices for $\mu$ and $p_r$.
\section{General solution to the field equations\label{sec:III}}

It is convenient to make the following choice for the energy
density
\begin{eqnarray}
\mu & = & \frac{j}{r^2}+k+\ell r^2 \label{muIII}
\end{eqnarray}
where $j$, $k$ and $\ell$ are constants. The roles of $j$, $k$ and
$\ell$ in the physics of the model are highlighted in examples
considered later. An advantage of this form for $\mu$ is that it
contains particular cases studied previously. Then (\ref{massFun})
yields the following expression for the mass function
\begin{eqnarray}
 m & = &
\frac{r}{2}\left(j+\frac{k}{3}r^2+\frac{\ell}{5}r^4\right)\label{mIII}
\end{eqnarray}
with the particular energy density (\ref{muIII}). Equation
\eqref{EFEs2:1} gives
\begin{eqnarray}
e^{-\lambda} & = & 1-j
-\frac{k}{3}r^2-\frac{\ell}{5}r^4\label{elambdaIII}
\end{eqnarray}
and the gravitational potential $\lambda$ has been determined.

With the help of (\ref{mIII}), we can write (\ref{EFEs2:2}) as
\begin{eqnarray}
\nu^\prime & = &
\frac{rp_r}{1-j-\frac{k}{3}r^2-\frac{\ell}{5}r^4}+\frac{j+\frac{k}{3}r^2+\frac{\ell}{5}r^4}{r\left(1-j
-\frac{k}{3}r^2-\frac{\ell}{5}r^4\right)}\nonumber \\
& = & \frac{r p_r}{1-j-\frac{k}{3}r^2-\frac{\ell}{5}r^4}
+\frac{j}{r\left(1-j\right)}\nonumber \\ & &
+\frac{\frac{k}{3}r+\frac{\ell}{5}r^3}{\left(1-j\right)\left(1-j
-\frac{k}{3}r^2-\frac{\ell}{5}r^4\right)} \label{nuprimeIII:2}
\end{eqnarray}
where we have used partial fractions. On integration,
\eqref{nuprimeIII:2} can be expressed as
\begin{eqnarray}
\nu & = & I_1 + \frac{j}{1-j}\ln r+\frac{1}{1-j}I_2+\ln B
\label{nuIII}
\end{eqnarray}
where $\ln B$ is a constant of integration and we have set
\begin{eqnarray*}
I_1 & = & \int \frac{r p_r}{1-j-\frac{k}{3}r^2-\frac{\ell}{5}r^4} \mbox{d}r \\ 
I_2 & = & \int\frac{\frac{k}{3}r+\frac{\ell}{5}r^3}{1-j
-\frac{k}{3}r^2-\frac{\ell}{5}r^4}\mbox{d}r 
\end{eqnarray*}
At this point we could choose a barotropic equation of state
$p_r=p_r(\mu)$. However this is an approach that we intend to
follow in future work. In this treatment we make a choice for the
radial pressure $p_r$ which is physically reasonable and is a
generalisation of earlier studies. We make the choice
\begin{eqnarray}
p_r & = &
\frac{C}{1-j}\left(1-j-\frac{k}{3}r^2-\frac{\ell}{5}r^4\right)\left(1-\frac{r^2}{R^2}\right)^n
\label{pr_choice}
\end{eqnarray}
When $j=\ell=0$, we obtain the radial pressure postulated by
Maharaj and Maartens \cite{MaharajMaartens}. For $j=0$, we regain
the radial pressure of Gokhroo and Mehra \cite{gokhroo}. The form
\eqref{pr_choice} for $p_r$ is physically reasonable because
$p_r>0$ in the interval $\left(0,R\right)$ for relevant choices of
the constants, $p_r=C$ at the centre $r=0$, $p_r=0$ at the
boundary $r=R$, and $p_r$ is continuous and well behaved in the
interval $\left[0,R\right]$.

The first integral $I_1$ simplifies to
\begin{eqnarray*}
I_1 & = &
\frac{C}{1-j}\int\left(1-\frac{r^2}{R^2}\right)^nr\mbox{d}r \\
& = &
-\frac{CR^2}{2(1-j)(n+1)}\left(1-\frac{r^2}{R^2}\right)^{n+1}
\end{eqnarray*}
for the choice of \eqref{pr_choice}. To evaluate the second
integral $I_2$ we need to consider two cases: $\ell=0$ and
$\ell\neq 0$.
\newpage
\textit{ Case I:} $\ell=0$

In this case the integration is straightforward and we obtain
\begin{eqnarray*}
I_2 & = & -\frac{1}{2}\ln\left\{1-j-\frac{k}{3}r^2\right\}
\end{eqnarray*}

\textit{ Case II:} $\ell\neq 0$

For this case we let
\[
u  =  r^2+\frac{5k}{6\ell}, \;\; q^2  =  1-j+\frac{5k^2}{36\ell}
\]
and obtain
\begin{eqnarray*}
I_2 & = & \int
\frac{\frac{\ell r}{5}\left(r^2+\frac{5k}{3\ell}\right)}{1-j+\frac{5k^2}{36\ell}-\frac{\ell}{5}\left(r^2+\frac{5k}{6\ell}\right)^2}\mbox{d}r\\
& = &
\frac{\ell}{10}\int\frac{u+\frac{5k}{6\ell}}{q^2-\frac{\ell}{5}u^2}\mbox{d}u\\
& = & \frac{\ell}{10}\left(
-\frac{5}{2\ell}\ln\left\{q^2-\frac{\ell}{5}u^2\right\}+\frac{5k}{6\ell}\left(\frac{\sqrt{5}}{q\sqrt{\ell}}\right)\tanh^{-1}\left\{\frac{u\sqrt{\ell}}{q\sqrt{5}}\right\}
\right)
\end{eqnarray*}
Hence we can collectively write for both \textit{Case I} and
\textit{Case II} that
\begin{eqnarray}
I_2 & = & \left\{\begin{array}{l}
-\frac{1}{2}\ln\left\{1-j-\frac{k}{3}r^2\right\}\mbox{,
for } \ell=0 \\ \\
-\frac{1}{4}\ln\left\{1-j+\frac{5k^2}{36\ell}-\frac{\ell}{5}\left(r^2+\frac{5k}{6\ell}\right)^2\right\}\\
+\left(\frac{5}{\ell}\right)^{\frac{1}{2}}\left(\frac{k}{12\sqrt{1-j+\frac{5k^2}{36\ell}}}\right)
\tanh^{-1}\left\{\left(\frac{\ell}{5}\right)^{\frac{1}{2}}\frac{r^2+\frac{5k}{6\ell}}{\sqrt{1-j+\frac{5k^2}{36\ell}}}\right\}\mbox{,
for } \ell\ne 0
\end{array}\right. \label{I2_Lcases}
\end{eqnarray}
The integrals $I_1$ and $I_2$ are given in terms of elementary
functions which helps in the physical analysis of the model.

On substituting $I_1$ in (\ref{nuIII}) we obtain
\begin{eqnarray}
e^\nu & = &
Br^{\frac{j}{1-j}}\exp\left\{\frac{I_2}{1-j}-\frac{CR^2}{2(1-j)(n+1)}\left(1-\frac{r^2}{R^2}\right)^{n+1}\right\}\label{e_nuIII}
\end{eqnarray}
for the gravitational potential $e^\nu$ where $I_2=I_2(r)$ has the
functional representation given above in (\ref{I2_Lcases}) for
$\ell=0$ and $\ell\ne 0$. To match the interior solution to the
Schwarzschild interior we require that $e^{\nu(R)}=1-2M/R$ which
implies that $B=R^{\frac{-j}{1-j}}\left(1-2M/R\right)\exp
\left\{-I_2(R)/(j-1)\right\}$. Finally the last field equation
(\ref{EFEs2:3}) gives the tangential pressure $p_\perp$:
\begin{eqnarray}
p_\perp & = &
p_r+\frac{C}{2\left(1-j\right)}\left(j-\frac{\ell}{5}r^4\right)\left(1-\frac{r^2}{R^2}\right)^{n}\nonumber\\
& & +\frac{r^2}{2}\left(1-j-\frac{k}{3}r^2-\frac{\ell}{5}r^4\right)^{-1}\nonumber\\
& &
\times\left\{\frac{C^2}{2(1-j)^2}\left(1-j-\frac{k}{3}r^2-\frac{\ell}{5}r^4\right)^2\left(1-\frac{r^2}{R^2}\right)^{2n}\right.\nonumber\\
& &
-\frac{2nC}{(1-j)R^2}\left(1-j-\frac{k}{3}r^2-\frac{\ell}{5}r^4\right)^2\left(1-\frac{r^2}{R^2}\right)^{n-1}\nonumber\\
& & \left. +\frac{1}{2r^2}\left(\frac{j}{r^2}+k+\ell
r^2\right)\left(j+\frac{k}{3}r^2+\frac{\ell}{5}r^4\right)\right\}\label{p_perp}
\end{eqnarray}
where we have used (\ref{muIII}), (\ref{pr_choice}) and
(\ref{e_nuIII}). The anisotropic factor $S(r)$ is given by
\begin{eqnarray}
S & = &
-\frac{C}{2\sqrt{3}\left(1-j\right)}\left(j-\frac{\ell}{5}r^4\right)\left(1-\frac{r^2}{R^2}\right)^{n}\nonumber\\
& & -\frac{r^2}{2\sqrt{3}}\left(1-j-\frac{k}{3}r^2-\frac{\ell}{5}r^4\right)^{-1}\nonumber\\
& &
\times\left\{\frac{C^2}{2(1-j)^2}\left(1-j-\frac{k}{3}r^2-\frac{\ell}{5}r^4\right)^2\left(1-\frac{r^2}{R^2}\right)^{2n}\right.\nonumber\\
& &
-\frac{2nC}{(1-j)R^2}\left(1-j-\frac{k}{3}r^2-\frac{\ell}{5}r^4\right)^2\left(1-\frac{r^2}{R^2}\right)^{n-1}\nonumber\\
& & \left. +\frac{1}{2r^2}\left(\frac{j}{r^2}+k+\ell
r^2\right)\left(j+\frac{k}{3}r^2+\frac{\ell}{5}r^4\right)\right\}
\label{anis_factor}
\end{eqnarray}
which follows from (\ref{pr_choice}) and (\ref{p_perp}).

Thus we have generated a new class of solutions to the Einstein
field equations (\ref{EFEs2}).  Collecting the various results
given above we can express the exact solution as
\begin{subequations}
\begin{eqnarray}
\mu & = & \frac{j}{r^2}+k+\ell r^2\label{XSols:1}\\
p_r & = &
\frac{C}{1-j}\left(1-j-\frac{k}{3}r^2-\frac{\ell}{5}r^4\right)\left(1-\frac{r^2}{R^2}\right)^n\label{XSols:2}\\
p_\perp & = &
p_r+\frac{C}{2\left(1-j\right)}\left(j-\frac{\ell}{5}r^4\right)\left(1-\frac{r^2}{R^2}\right)^{n}
+\frac{r^2}{2}\left(1-j-\frac{k}{3}r^2-\frac{\ell}{5}r^4\right)^{-1}\nonumber\\
& &
\times\left\{\frac{C^2}{2(1-j)^2}\left(1-j-\frac{k}{3}r^2-\frac{\ell}{5}r^4\right)^2\left(1-\frac{r^2}{R^2}\right)^{2n}\right.\nonumber\\
& &
-\frac{2nC}{(1-j)R^2}\left(1-j-\frac{k}{3}r^2-\frac{\ell}{5}r^4\right)^2\left(1-\frac{r^2}{R^2}\right)^{n-1}\nonumber\\
& & \left. +\frac{1}{2r^2}\left(\frac{j}{r^2}+k+\ell
r^2\right)\left(j+\frac{k}{3}r^2+\frac{\ell}{5}r^4\right)\right\}\label{XSols:3}\\
e^\nu & = & Br^{\frac{j}{1-j}}\exp\left\{\frac{I_2}{1-j}-\frac{CR^2}{2(1-j)(n+1)}\left(1-\frac{r^2}{R^2}\right)^{n+1}\right\}\label{XSols:4}\\
e^{\lambda} & = & \frac{1}{1-j
-\frac{k}{3}r^2-\frac{\ell}{5}r^4}\label{XSols:5}
\end{eqnarray}\label{XSols}
\end{subequations}
where $I_2$ (given in (\ref{I2_Lcases})) contains the two cases,
$\ell=0$ and $\ell\ne 0$. The exact solution (\ref{XSols})
represents the interior of an anisotropic star corresponding to
the energy density $\mu=j/r^2+k+\ell r^2$. Clearly other choices
for $\mu$ and $p_r$ will yield new solutions to the field
equations; however these choices may not correspond to realistic
matter or the integrals $I_1$ and $I_2$ may not  be expressible in
closed form. This solution does not have a barotropic equation of
state $p_r=p_r(\mu)$. To obtain a model with an equation of state
we need to specify this explicitly when evaluating the integral
$I_1$. An equation of state is a desirable physical feature which
we hope to incorporate in future models. Note that the
cosmological constant is absent from our model. This quantity can
be easily included by adding a constant to the energy density and
the pressure function.

\section{Special cases\label{sec:IV}}

We consider some special cases contained in the new class of
solution presented in section \ref{sec:III}; two of these cases
lead to particular models that have been studied previously.

\textit{Solution I:} $j=0$

In this case the energy density is given by
\begin{eqnarray*}
\mu & = & k+\ell r^2\label{XSolsGM:1}\end{eqnarray*} and the line
element has the form
\begin{eqnarray}
\mbox{d}s^2 & =&
-\left(B\exp\left\{I_2-\frac{CR^2}{2(n+1)}\left(1-\frac{r^2}{R^2}\right)^{n+1}\right\}\right)\mbox{d}t^2\nonumber\\
& &
+\left(1-\frac{k}{3}r^2-\frac{\ell}{5}r^4\right)^{-1}\mbox{d}r^2+r^2\left(\mbox{d}\theta^2+\sin^2\theta\mbox{d}\phi^2\right)
\label{line:genGM}
\end{eqnarray}
The particular solution \eqref{line:genGM} was found by Gokhroo
and Mehra \cite{gokhroo}. Their solution is regained when we set $
k  = \rho_0, \;\; \ell  =  -\rho_0 K/a^2 $. Note that if we
require $\mu^\prime<0$ then the constant $\ell<0$ for a
monotonically decreasing energy density as we approach the
boundary $r=R$ from the centre.

\textit{Solution II:} $j=\ell=0$

For this case the energy density
\begin{eqnarray*}
\mu & = & k
\end{eqnarray*}
is a constant. The line element has the representation
\begin{eqnarray}
\mbox{d}s^2 & =&
-\left(B\exp\left\{-\frac{1}{2}\ln\left(1-\frac{k}{3}r^2\right)-\frac{CR^2}{2(n+1)}\left(1-\frac{r^2}{R^2}\right)^{n+1}\right\}\right)\mbox{d}t^2\nonumber\\
& &
+\left(1-\frac{k}{3}r^2\right)^{-1}\mbox{d}r^2+r^2\left(\mbox{d}\theta^2+\sin^2\theta\mbox{d}\phi^2\right)
\label{line:genMM}
\end{eqnarray}
The particular solution \eqref{line:genMM} was found by Maharaj
and Maartens \cite{MaharajMaartens}. Their solution is regained
when we let $ k = 6M/R^3 $. Since $\mu$ is constant we may
interpret this solution as an anisotropic generalisation of the
incompressible Schwarzschild interior sphere; however note that
the anisotropy factor $S(r)\ne 0$ everywhere except at the centre
$r=0$.

\textit{Solution III:} $k=\ell=0$

In this case the energy density has the form
\begin{eqnarray*}
\mu & = & \frac{j}{r^2}
\end{eqnarray*}
The line element is given by
\begin{eqnarray}
\mbox{d}s^2 & =&
-Br^{\frac{j}{1-j}} \nonumber\\
& &\times\exp\left\{-\frac{1}{2}\ln\left(1-j\right)-\frac{CR^2}{2(1-j)(n+1)}\left(1-\frac{r^2}{R^2}\right)^{n+1}\right\}\mbox{d}t^2\nonumber\\
& &
+\left(1-j\right)^{-1}\mbox{d}r^2+r^2\left(\mbox{d}\theta^2+\sin^2\theta\mbox{d}\phi^2\right)
\label{line:genISO}
\end{eqnarray}
Even though \eqref{line:genISO} has a very simple form, we believe
that it is a new anisotropic solution to the Einstein field
equations and has not been published before. Since $\mu\propto
r^{-2}$ we may relate \eqref{line:genISO} to the results of other
treatments. Dev and Gleiser \cite{DevGleiser}, Herrera and Santos
\cite{HerreraSantos2} and Petri \cite{Petri} found solutions to
the anisotropic Einstein field equations involving $\mu\propto
r^{-2}$. In each of these papers a different set of assumptions to
that utilised in this paper was used; in our treatment we have
chosen a form for the radial pressure $p_r$. Therefore their
solutions are necessarily different from \eqref{line:genISO} for
the corresponding energy density choice $\mu\propto r^{-2}$.
\section{Physical Conditions and Analysis\label{sec:V}}

One of the original reasons for studying anisotropic matter was to
generate models that permit redshifts higher than the critical
redshift $z_c$ of isotropic matter \cite{BowersLiang}.
Observational results indicate that certain isolated objects have
redshifts higher than $z_c$. The surface redshift is given by
\begin{eqnarray*}
z & = & \left(1-\frac{2M}{R}\right)^{-\frac{1}{2}}-1
\end{eqnarray*}
The critical redshift $z_c=2$ is the limiting value for the
perfect fluid spheres, and is attained when $2M/R=8/9$
\cite{buchdahl}. For the range of values falling in the interval
$8/9<2M/R<1$ the redshift is greater than $z_c$; this phenomenon
may be explained by allowing for anisotropy. For values of $2M/R$
close to unity, the surface redshift becomes infinitely large. The
feasibility of higher redshifts for anisotropic matter, in both
Newtonian and relativistic models, was firmly established by Bondi
\cite{bondi}. It is interesting to note that Bondi , Binney and
Tremaine \cite{binneytremaine}, Cuddeford \cite{cuddeford} and
Michie \cite{michie} emphasise the significance of anisotropies in
stellar clusters and galaxies, in addition to individual stars.

The gravitational potential $e^{\lambda}$ is finite at the centre
$r=0$ and at the boundary $r=R$. The function $e^\lambda$ is well
behaved in the interior of the relativistic star. The
gravitational potential $e^\nu$ is continuous and well behaved in
the interior and finite at the boundary of the star $r=R$. There
is a singularity at the centre $r=0$ in the potential $e^\nu$. The
singularity in $e^\nu$ is removable for a specific choice of
parameter values. This singularity is eliminated by setting $j=0$
which corresponds to the solution of Gokhroo and Mehra
\cite{gokhroo}.

The energy density \eqref{muIII} chosen describes relativistic
stars as we demonstrate later. The form of $\mu\mbox{ }\propto
r^{-2}\mbox{ }(k=\ell=0)$ is usually used in domains where it is
not possible to use a single equation of state; particularly where
the origin is excluded, like a body with a constant density core
and matter density distribution around the core going like
$r^{-2}$ \cite{DevGleiser,SharmaMukherjee2002}. It is interesting
to observe that the $r^{-2}$ profile in the energy density also
arises in isothermal spheres in Newtonian configurations that
correspond to a Maxwell-Boltzmann gas in galactic systems
\cite{SaslawMaharaj}. Densities with $j\neq 0$ and $k\neq 0$ are
also physically reasonable. For example, Misner and Zapolsky
\cite{MisnerZapolsky} propose that the term $jr^{-2}$ models the
physical configuration of a relativistic Fermi gas for some
particular value of the parameter $j$. Another example is due to
Dev and Gleiser \cite{DevGleiser} who suggest that for some
particular value of $j$ and $k\ne 0$  the energy density function
$jr^{-2}+k$ describes a relativistic Fermi gas core immersed in a
constant density background.

The radial pressure $p_r$ is continuous and well behaved in the
interior of the star. Also $p_r>0$ in the interval $(0,R)$,
regular at the centre $\left(p_r(r=0)=C\right)$, and vanishes at
the boundary $\left(p_r(r=R)=0\right)$. The tangential pressure
$p_\perp$ has a singularity at the centre, but is otherwise well
behaved throughout the interior of the star and finite at the
boundary. The singularity in $p_\perp$ may be eliminated by
suitable particular choice of parameter values. In general the
tangential pressure is not zero at the boundary of the star
$\left(p_\perp(r=R)\ne 0\right)$ which is different from the
radial pressure $\left(p_r(r=R)= 0\right)$. It is also important
to observe that the magnitude of the stress tensor
\begin{eqnarray*}
S & = & \frac{1}{\sqrt{3}}\left(p_r-p_\perp\right)
\end{eqnarray*}
is a nonzero function in general. Hence this class of solutions is
generally anisotropic and does not have an isotropic limit (the
isotropic limit results when we set particular values for the
constants in our ansatz). It is not possible to eliminate $S$ and
obtain an isotropic counterpart. This means that the model remains
anisotropic. An analogous situation rises in Einstein-Maxwell
solutions modelling charged relativistic stars in which the
electric field is always present. An example of such a charged
star is given by Hansraj \cite{hansraj}.

\begin{figure}[thb]
\vspace{1.5in} \includegraphics{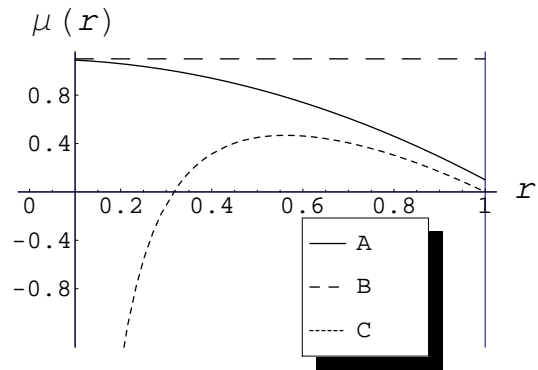}
\caption{\label{muplots}Energy density $\mu(r)$ plots}
\end{figure}

\begin{figure}[htb]
\vspace{1.5in} \includegraphics{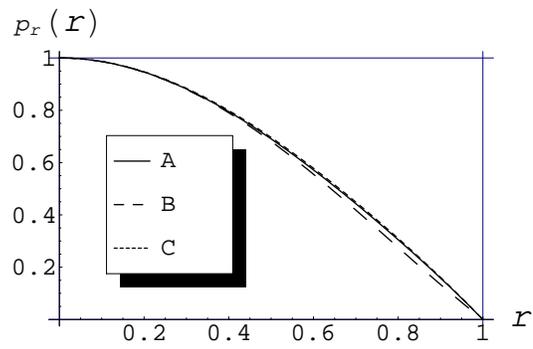}
\caption{\label{prplots}Radial pressure $p_r(r)$ plots}
\end{figure}

\begin{figure}[hbt]
\vspace{1.4in} \includegraphics{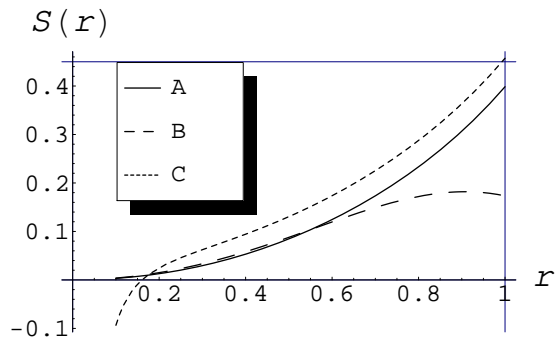}
\caption{\label{Splots}Anisotropy factor $S(r)$ plots}
\end{figure}

Figures \ref{muplots}, \ref{prplots}, and \ref{Splots} are
illustrations of the behaviour of the energy density $\mu$, the
radial pressure $p_r$, and the anisotropy factor $S$ respectively,
for particular chosen values of the constants in the exact
solution \eqref{XSols}. The radial distance is over the interval
$0\leq r \leq 1$ and the boundary of the star has been normalised
to be $r=R=1$. Note that Plot A corresponds to $j=0$ case
(\emph{Solution I}), Plot B corresponds to $j=\ell=0$
(\emph{Solution II}) and Plot C corresponds to the general
solution \eqref{XSols} where $j\ne 0$, $k\ne 0$, and $\ell\ne 0$.
In Figure \ref{muplots} for $\mu$, Plots A and B are continuous
throughout the interval $0\leq r \leq 1$; however Plot C indicates
unphysical behaviour as we approach the centre. This undesirable
feature in Plot C arises because $j\ne 0$ and indicates that
another solution has to be utilised around the centre in a
core-envelope model \cite{DevGleiser,HerreraSantos2,Petri}. In
Figure \ref{prplots} for $p_r$, the radial pressure is
monotonically decreasing from the centre to the boundary for all
Plots A, B, and C. In Figure \ref{Splots} for $S$, Plots A and B
are continuous throughout the interval $0\leq r \leq 1$; however
Plot C indicates a singularity as we approach $r=0$. We suspect
that this singularity in Plot C is related to the fact that $j\ne
0$. We observe that the gradient of $S$ is greatest for Plot C,
corresponding the case for $j\ne 0$, $k\ne 0$, and $\ell\ne 0$, as
the boundary is approached. Hence $S(R)$ has the largest value at
the boundary for the general solution \eqref{XSols} in this case.
The behaviour of $S$ outside the centre is likely to correspond to
physically reasonable anisotropic matter: Plot B has a profile
similar to the behaviour of the anisotropic boson stars studied by
Dev and Gleiser \cite{DevGleiser}. However the general solution
obtained by Dev and Gleiser \cite{DevGleiser} for the choice
\mbox{$\mu=jr^{-2}+k$} is given in terms of hypergeometric
functions. Our corresponding solution has the advantage of being
expressed in terms of elementary functions.

\newpage
\vspace{.1in}
\vspace{.1in}
\newpage
\section{Discussion \label{sec:VI}}

It is possible to demonstrate that the solutions found can be
utilised to discuss the structure of neutron stars and
quasi-stellar objects. We write the surface density in the
particular form $\mu_s=\mu_0(1-\tilde{j}-\tilde{\ell})$ where
$\mu_0$ is the central density and the constants $j$, $k$ and
$\ell$ have been chosen so that the energy density can be easily
expressed in c.g.s. units. Now consider a neutron star of radius
10km and surface density of $2\times 10^{14}\;\mbox{gcm}^{-3}$.
Then the parameters $\mu_0$, $\mu_0R^2$, $2M/R$, surface redshift
$z=\left(1-2M/R\right)^{-1/2}-1$, and mass $M$ in terms of the
solar masses $M_\odot$ can be calculated. We choose values of
$\tilde{j}$ and $\tilde{\ell}$ so that comparison with Gokhroo and
Mehra \cite{gokhroo} is facilitated. The results are given in
Table \ref{table}. In this category of results the surface
redshifts range up to $0.566$, and masses extend to $2.00M_\odot$.
This range of values is consistent with the results of Gokhroo and
Mehra \cite{gokhroo}. Hence our solutions yield values for surface
redshifts and masses that correspond to realistic stellar sources
such as Her X-1 and Vela X-1. Clearly higher values for $z$ and
$M$ can be generated by adjusting $\tilde{j}$ and $\tilde{\ell}$.

\begin{table}[htb]
\begin{center}
\begin{tabular}{|c|c|c|c|c|c|c|} \hline \hline
$\tilde{j}$  & $\tilde{\ell}$ & $\mu_0\times10^{14}$ & $\mu_0R^2$ & $2M/R$ &$z$&$M(M_\odot)$\\
\hline \hline
0 & 0 & 1.48 & .015 & .124 &.068 &.42\\
\hline  .001&  .1& 1.65&.017&.130&.072&.44\\ \hline  .002&.2 &1.86 &.019&.136&.076&.46 \\
\hline  .003&.3  &2.13 &.021&.145&.081&.49 \\\hline .004&.4  &2.49 &.025&.156&.089&.53 \\
\hline .005&.5 &3.00 &.030&.172&.099&.58
\\ \hline  .006&.6  &3.77 &.038&.196&.116&.66
\\ \hline  .007&.7  &5.07 &.051&.237&.145&.80
\\
\hline  .008&.8  &7.73 &.077&.321&.214&1.09
\\
\hline  .009&.9  &16.32 &.163&.592&.566&2.00
\\
\hline\hline
\end{tabular}
\caption{\label{table} Densities and redshifts for neutron stars}
\end{center}
\end{table}

In this paper we have found a new class of solutions to the
Einstein field equations for an anisotropic matter distribution
utilising the algorithm of Maharaj and Maartens
\cite{MaharajMaartens}. These solutions correspond to the energy
density $\mu=jr^{-2}+k+\ell r^2$ and contain particular solutions
found previously. We note that the term containing $jr^{-2}$ is
physically important and arises in a number of applications
\cite{DevGleiser,DevGleiser2,MisnerZapolsky,SharmaMukherjee2002}.
Our results indicate that anisotropic solutions for the physically
reasonable energy density $\mu\propto r^{-2}$ can be generated
with the simple solution generating mechanism of Maharaj and
Maartens \cite{MaharajMaartens}. Our ongoing investigations
indicate that a general class of anisotropic models are possible,
for different choices of $\mu$, such that the desired limit
$\mu\propto r^{-2}$ is regained as we approach the boundary. This
work is in preparation. Observe from \eqref{anis_factor} that the
anisotropy factor $S$ is nonzero in general in the interior of the
star. This means that the exact solution \eqref{XSols} remains
anisotropic and does not have an isotropic limit. We would need to
use another approach of integrating the anisotropic Einstein field
equations than the algorithm used in this paper, if an isotropic
limit is to be contained in the stellar model.

\section*{Acknowledgements}
MC is grateful to the University of KwaZulu-Natal for a
scholarship. MC and SDM thank the National Research Foundation for
financial support. We are grateful to the referees for their input
which has substantially improved the paper.

%

%
%
%
\end{document}